\documentclass[journal]{IEEEtran}
\usepackage{tikz}
\usetikzlibrary{shapes,arrows}

\pdfoutput=1

\usepackage{subcaption}
\ifCLASSINFOpdf
\else
\fi
\usepackage{amsmath}
\usepackage{multirow}
\DeclareMathOperator*{\argmin}{arg\,min}

\begin{document}
%
\title{Improved Frequency Modulation Features for Multichannel Distant Speech Recognition}
%
%
%

\author{\IEEEauthorblockN{Isidoros Rodomagoulakis and Petros Maragos\thanks{This research work was supported by the EU under the project I-SUPPORT with grant H2020-643666.}}\\
\IEEEauthorblockA{School of ECE, National Technical University of Athens,
15773 Athens, Greece\\
\{irodoma,maragos\}@cs.ntua.gr}
}
\maketitle

\begin{abstract}
Frequency modulation features capture the fine structure of speech formants that constitute beneficial and supplementary to the traditional energy-based cepstral features.  
Improvements have been demonstrated mainly in GMM-HMM systems for small and large vocabulary tasks. 
Yet, they have limited applications in  DNN-HMM systems and Distant Speech Recognition (DSR) tasks. 
Herein, we elaborate on their integration within state-of-the-art front-end schemes that include post-processing of MFCCs resulting in discriminant and speaker-adapted features of large temporal contexts. 
We explore  1)  multichannel demodulation schemes for multi-microphone setups, 2) richer descriptors of frequency modulations, and 3) feature transformation and combination via hierarchical deep networks. 
We present results for  tandem and hybrid recognition with GMM and DNN acoustic models, respectively. 
The improved modulation features are combined efficiently with MFCCs yielding modest and consistent improvements  in  multichannel distant speech recognition tasks on  reverberant and noisy  environments, where recognition rates are far from human performance.

\end{abstract}

\begin{IEEEkeywords}
Frequency modulation features, Demodulation, Deep bottleneck features, Distant speech recognition
\end{IEEEkeywords}

%
\IEEEpeerreviewmaketitle

\section{Introduction}
Modulation features stemming from the AM-FM speech model were originally conceived for ASR~\cite{Dimitriadis2005} as capturing the second-order non-linear structure of speech formants, providing complementary information to the traditional energy-based cepstral features (e.g., MFCCs and PLPs).
Their fusion presents robustness in noisy and mismatched conditions as indicated in recent works~\cite{dimitriadis2015use,mitra2014evaluating}. However, only a few works~\cite{Rodomagoulakis2013a,mitra2015improving} examine their performance in DSR tasks with reverberation.  Recently, bottleneck Multilayer Perceptons (MLPs)  have been proposed in~\cite{dimitriadis2015use} to combine frequency micro-modulation features with PLPs using network's non-linear transformations instead of Linear Discriminant Analysis (LDA) which is suboptimal for non-Gaussian  features. Following the tandem approach~\cite{hermansky2000tandem}, improved and deeper nets were proposed in~\cite{sainath2012auto,yu2011improved}, while hierarchical architectures~\cite{tuske2013deep} were beneficial for feature combination.

 Deep Neural Networks (DNNs) have resulted in innovative ways to improve feature extraction and acoustic modeling in speech recognition~\cite{yu2014automatic}. Recently, end-to-end systems~\cite{hannun2014deep} have been developed to combine all recognition stages into Recurrent Neural Networks (RNNs) of long memory in order to transform unsegmented sequences of raw speech signals into sequences of phone labels, outperforming in many cases the hybrid DNN-HMM state-of-the-art systems. However,  they require  large amounts of data and processing capacity while poor performance persists due to high levels of noise and reverberation in many DSR scenarios~\cite{kinoshita2016summary, harper2015automatic} commonly found in modern applications.  
 
Although DNNs can learn many types of variation depending on the training data, they can be  sensitive to data mismatches, while feature transformations learned in a data-driven way may not generalize well for out-of-domain  data. Model adaptation with regularization mechanisms~\cite{yu2013kl} and iVector based adaptation~\cite{saon2013speaker} have been proposed for coping with unseen acoustic data. However, robust acoustic features are typically used to improve acoustic models when dealing with noisy and channel-degraded acoustic data. A comprehensive survey on robust feature extraction strategies and features for DNN-based recognition can be found in \cite{mitra2017robust}.

Multi-microphone setups  with array processing~\cite{brandstein2013microphone} offer flexibility on multi-source and noisy acoustic scenes by capturing the spatial diversity of speech and non-speech sources, allowing more sophisticated front-ends with channel combination~\cite{liu2014using}, beamforming~\cite{xiao2016deep} and speech enhancement~\cite{delcroix2017multichannel}, which were recently revised and solved with DNNs. However, the most significant improvements have been achieved with multi-style training on multichannel data~\cite{swietojanski2013hybrid,ko2017study}, while incorporating deep learning in traditional array processing methods is still under investigation.


Our goal in this work is to increase the robustness of frequency modulation features in noise and reverberation in order to combine them efficiently with standard MFCC-based frond-ends for state-of-the-art speech recognition with GMM and DNN acoustic models. First, we propose a Multichannel, Multiband Demodulation (MMD) scheme that utilizes the noise diversity across microphone array signals aiming at  improved demodulation of speech resonances and more accurate estimations of instantaneous modulations~\cite{tsiakoulis2013instantaneous}. Secondly, we explore richer representations of the estimated modulations either by applying signal compression on the raw signals, or by transforming mid-duration temporal contexts of their first-order statistics into hierarchical deep bottleneck networks, which are able to combine both non-linear transformation and fusion of heterogeneous features. Finally, we incorporate the proposed features combined with MFCCs in standard recognition recipes leveraging multi-style training and  beamforming. Experiments are conducted in simulated and real data with strong background noise and reverberation.

Section~\ref{sec:demodulation} presents  the proposed MMD approach with indicative results on the demodulation of speech phonemes;  Section~\ref{sec:features} describes the extraction of frequency modulation features along with the proposed hierarchical bottleneck DNN scheme; The  experimental  framework and the employed DSR corpora  are described in Section~\ref{sec:experiments}, while Sections ~\ref{sec:results} and ~\ref{sec:conclusions}  present the obtained results and the conclusions of the paper.

\section{Multichannel, Multiband Demodulation}
\label{sec:demodulation}
The proposed demodulation scheme exploits the spatial diversity of noise $u_m(t)$ exhibited across the $M$ recordings 
\begin{equation}
y_m(t) = s(t) + u_m(t),\ m=1,\dots, M
\end{equation}
of a microphone array capturing the clean source speech signal $s(t)$ in the continuous time domain $t$. 
 Note that reverberation effects and time alignment issues between $y_m(t)$ are not taken into account in the following analysis. 
 The recordings can be decomposed  into $N$ frequency bands  for the derivation of their bandpass components $y_{mk}(t),\;k=1,\dots, N$, which correspond to speech resonances. 
The $k$th resonance of the recording $y_m$  can be modeled  by an AM--FM signal as
\begin{equation}
y_{mk}(\,t)\;=\;a_{mk}(\,t)\cos\big(\int_{0}^{\,t}\!\!{\omega_{mk}(\tau)\:d\tau}\big)
\;,\;\;
\label{eq:resonance}
\end{equation}
where $a_{mk}(\,t)$ and $\omega_{mk}(\,t)$ are  its instantaneous amplitudes and angular frequencies.
We can track the energy of the source that produced the signal via the Teager-Kaiser energy operator (TEO)~\cite{kaiser1990simple}
\begin{equation}
\Psi[x(t)] = [\dot{x}]^2 - x(t)\ddot{x}(t)
\end{equation}
where $\dot{x} = dx(t)/dt$.
The TEO is the basic ingredient of the Energy Separation Algorithm (ESA)~\cite{Maragos1993} to demodulate the bandpass speech signals into instantaneous amplitude and frequency components.
Bandlimited speech components are obtained by decomposing $y_m(t)$ with a Mel-spaced Gabor filterbank $\{g_k(t)\}$:
\begin{equation}
y_{mk}(t) = y_m(t) * g_k(t),\ k=0,\dots, N-1
\end{equation}
where $g_k(t)$ corresponds to the impulse response of  the bandbass Gabor  filter over band $k$.
Given the correlated $k$th bandpass signals from any two  microphones  $m$ and $\ell$, their interaction can be described by the cross-Teager energy operator~\cite{kaiser1993some,maragos1995higher}:
\begin{equation}
\Psi_c[y_{mk},y_{\ell k}](t) = \dot{y}_{mk}(t)\dot{y}_{\ell k}(t) - y_{mk}(t)\ddot{y}_{\ell k} (t)
\end{equation}
which in general measures the relative rate of change between two oscillators.
As discussed in~\cite{lefkimmiatis2008multisensor} and~\cite{rodomagoulakis2017improvement}, two useful properties of the operator can be derived: 

1) On averaging, noise $u_m(t)$ contributes as an additive error term to the Teager energy $\Psi[s_k]$ of the $k$th resonance of the source signal: 
\begin{equation}
\mathcal{E}\{\Psi_c[y_{mk},y_{\ell k}]\}=\mathcal{E}\{\Psi[s_k]\}+ \textnormal{error}
\end{equation}
The above stands assuming that the additive noise component is a zero mean, wide-sense stationary (WSS) Gaussian random process. Consequently, 
the energy with the minimum average
\begin{equation}
\Psi^{\textnormal{min}}_c(k) = \Psi_c[y_{\hat{m}k},y_{\hat{\ell}k}]
\end{equation}
which is formed by microphones $(\hat{m},\hat{\ell})$, is expected to lie closer to $\Psi[s_k(t)]$. 

2) Instead of searching  $(\hat{m},\hat{\ell})$  among all pairs of microphones, which is computationally intensive\footnote{$2\cdot{{M}\choose{2}}$ computations are needed for each band because $\Psi_c[y_{mk},y_{\ell k}]\}\neq\Psi_c[y_{mk},y_{\ell k}]$}, it suffices to search between microphones $\bar{m}$ and $\bar{\ell}$ having the 1st and 2nd smallest average Teager energies:

\begin{equation}
(\hat{m},\hat{\ell}) = \argmin_{\bar{m},\bar{\ell}}\bigl(\mathcal{E}\{\Psi_c[y_{\bar{m}k},y_{\bar{\ell}k}]\},\, \mathcal{E}\{\Psi_c[y_{\bar{\ell}k},y_{\bar{m}k}]\}\bigr)
\end{equation}
Based on the above, we track  $\Psi^{\textnormal{min}}_c(k)$ in medium-duration frames in order to obtain an energy signal which is less affected by noise.
Then, we modify ESA, 
where instead of using single-channel energies $\Psi[y_{mk}]$,  we estimate the instantaneous amplitudes $a_{k}(t)$  and angular frequencies $\omega_{k}(t)$ using the denoised cross-channel energies as:\footnote{The microphone index $m$ is removed from now on as an indication of multi-microphone estimation.}
\begin{eqnarray}
\omega_{k}(t) \approx \sqrt{\Psi_c[\dot{y}_{\hat{m}k},\dot{y}_{\hat{\ell}k}] /\Psi_c[y_{\hat{m}k},y_{\hat{\ell}k}]} \\
\alpha_{k}(t) \approx \Psi_c[y_{\hat{m}k},y_{\hat{\ell}k}]/\sqrt{\Psi_c[\dot{y}_{\hat{m}k},\dot{y}_{\hat{\ell}k}] }
\end{eqnarray}
For computanional efficiency and smoother estimations, we compute cross-Teager energies by including bandpass filtering with Gabor filters $g_k(t)$ within the cross-Teager operator:
\begin{eqnarray}
\Psi_c[y_{\hat{m}k},y_{\hat{\ell}}] = (y_{\hat{m}} * \dot{g}_k)  (y_{\hat{\ell}} * \dot{g}_k) - (y_{\hat{m}} * g_k)  (y_{\hat{\ell}} * \ddot{g}_k) \\
\Psi_c[\dot{y}_{\hat{m}k},\dot{y}_{\hat{\ell}k}] = (y_{\hat{m}} * \ddot{g}_k) (y_{\hat{\ell}} * \ddot{g}_k) - (y_{\hat{m}} * \dot{g}_k)  (y_{\hat{\ell}} * \dddot{g}_k) 
\end{eqnarray}


\subsection{Analysis on TIMIT corpus}

\begin{figure*}[t!]
\centering
\begin{subfigure}[b]{0.24\linewidth}
\includegraphics[width=\columnwidth]{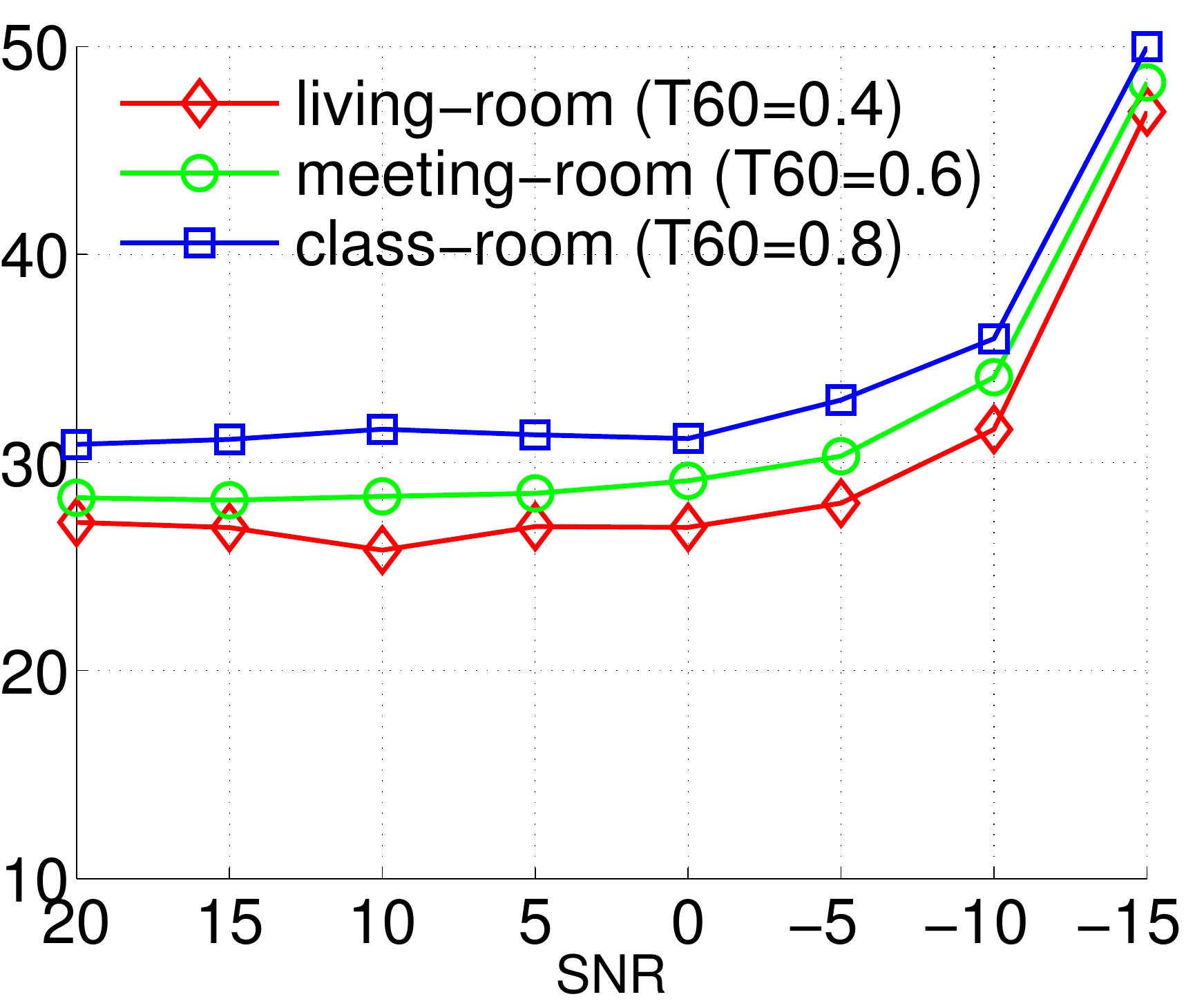}
\caption{vowels}
\end{subfigure}
\begin{subfigure}[b]{0.24\linewidth}
\includegraphics[width=\columnwidth]{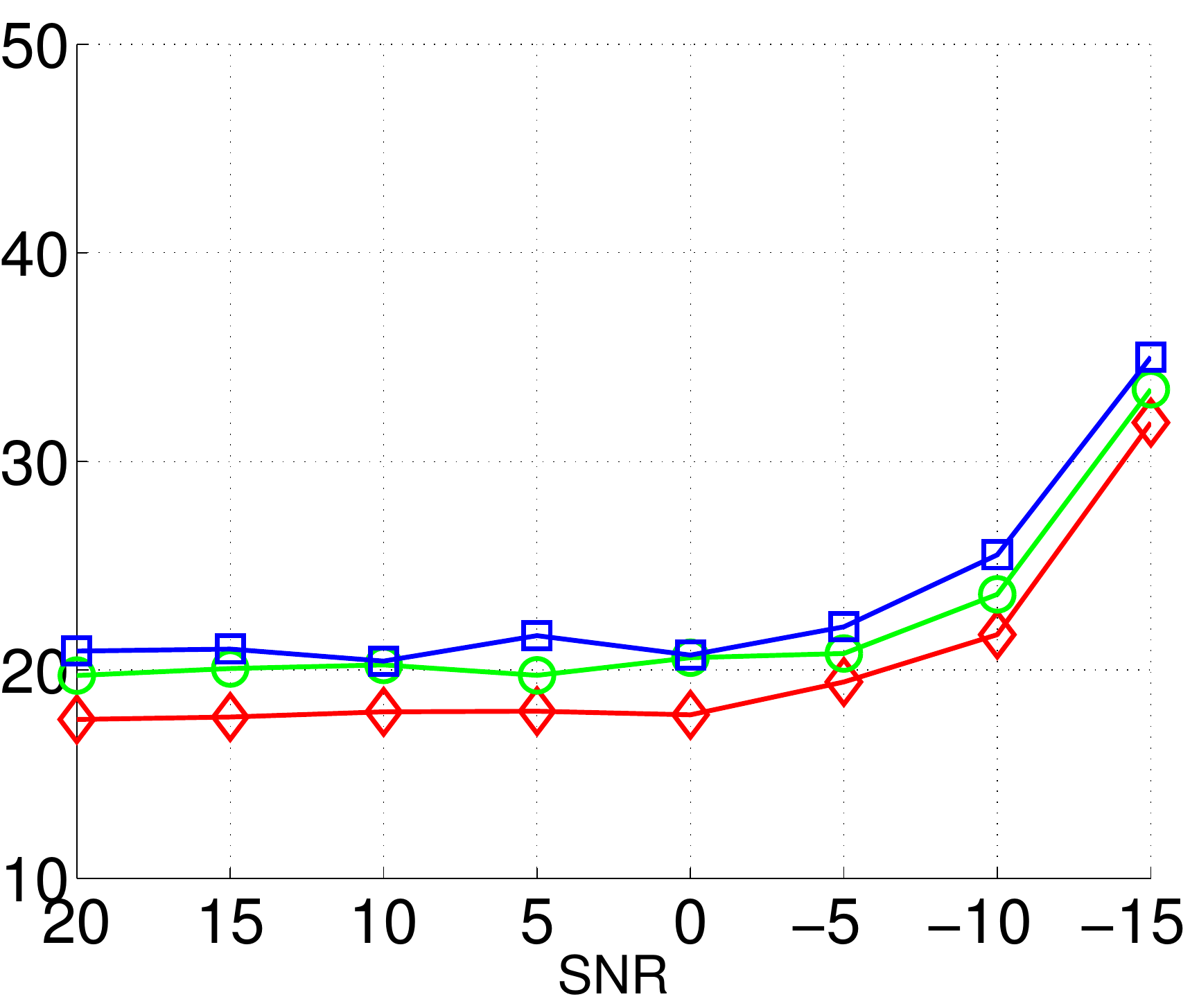}
\caption{nasals}
\end{subfigure}
\begin{subfigure}[b]{0.24\linewidth}
\includegraphics[width=\columnwidth]{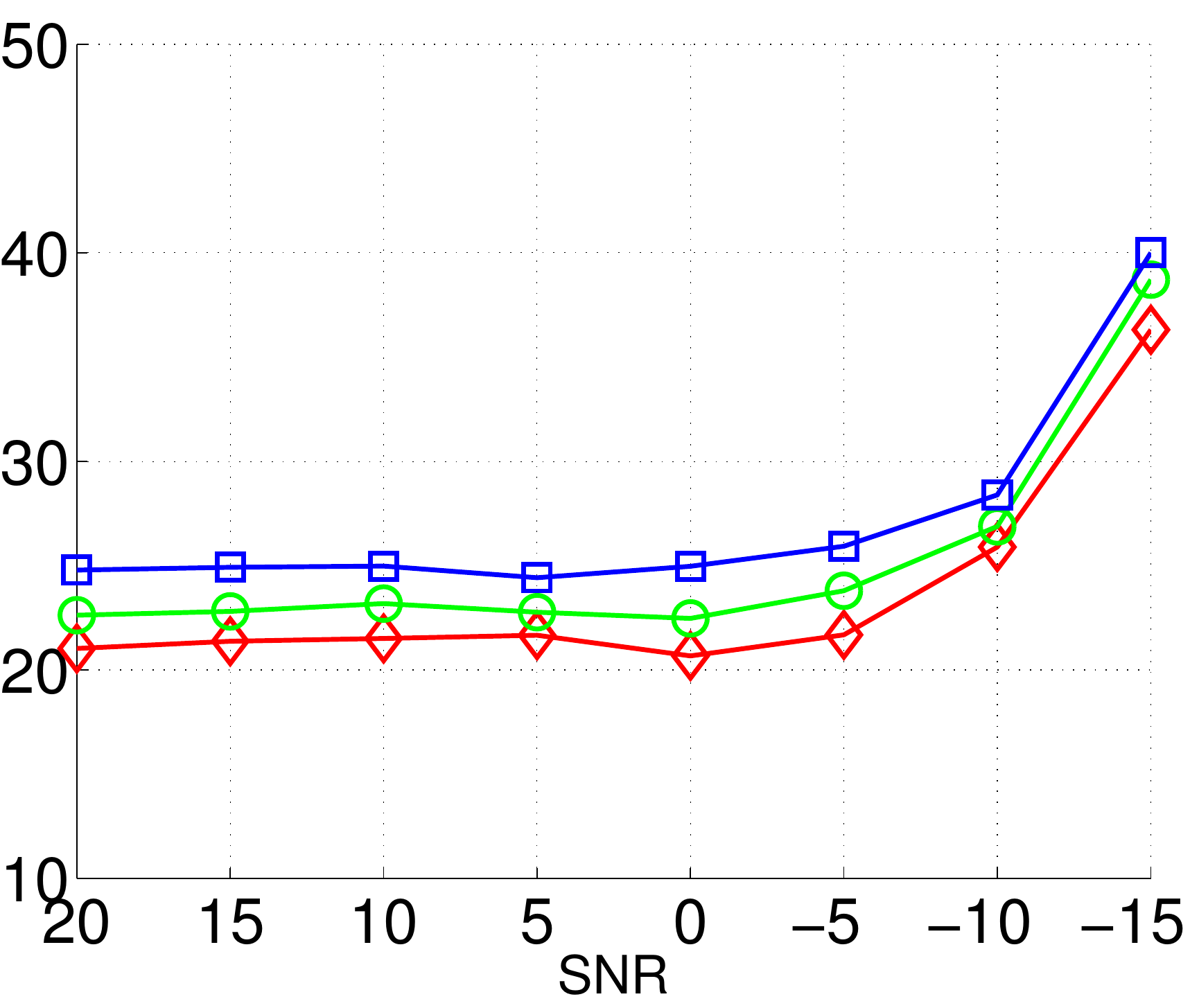}
\caption{stops}
\end{subfigure}
\begin{subfigure}[b]{0.24\linewidth}
\includegraphics[width=\columnwidth]{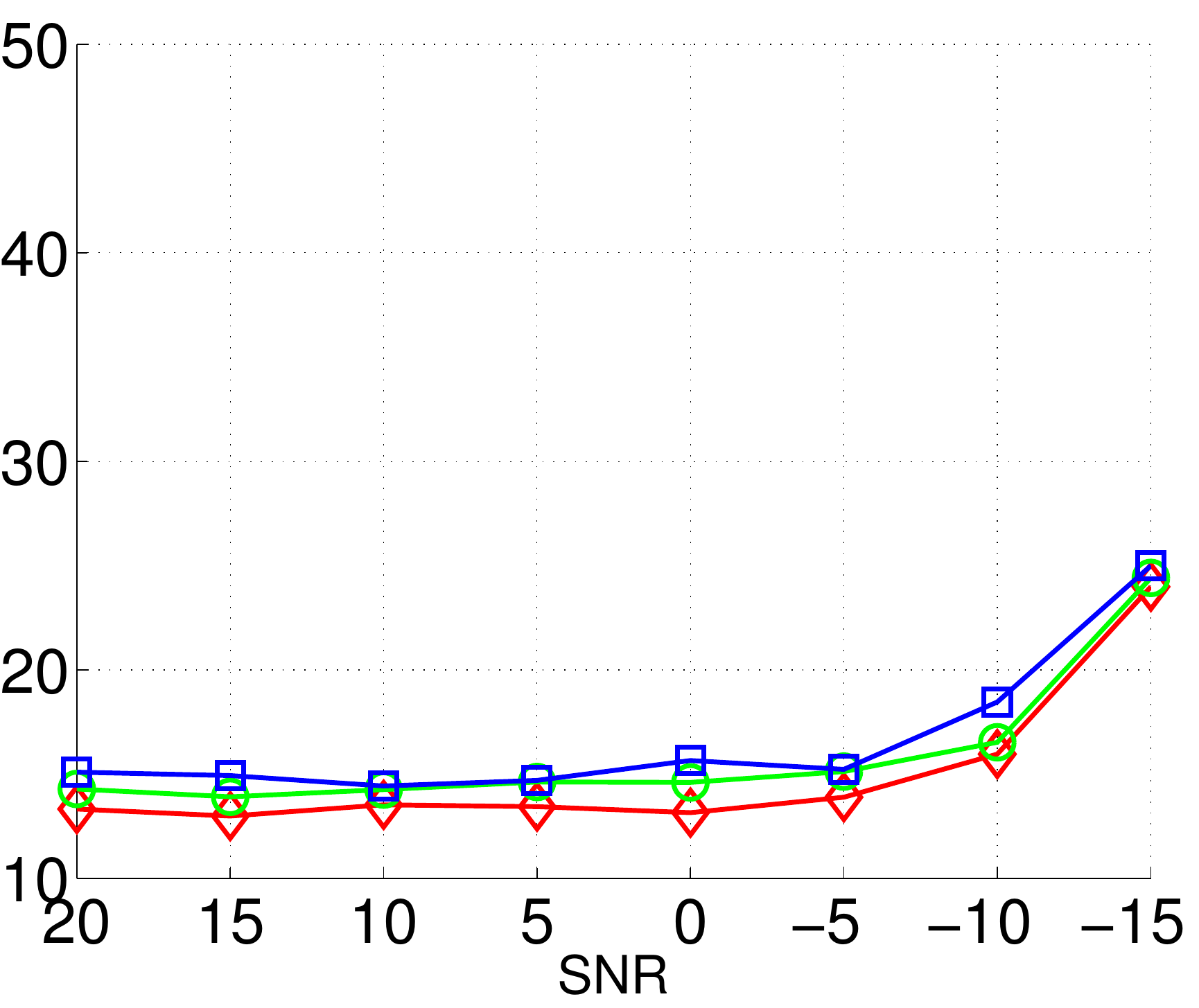}
\caption{fricatives}
\end{subfigure}
\caption{Relative reduction (\%) of frequency demodulation RMS error per  phoneme category achieved by the proposed MMD approach compared to single-channel Gabor-ESA demodulation.}
\label{fig:timit}
\end{figure*}

The robustness of the proposed MMD method is tested on simulations of noisy and far-field speech after distorting the TIMIT corpus. MMD is compared to the single-channel Gabor-ESA~\cite{Dimitriadis2006} in terms of frequency demodulation RMS error which is computed across bands and between the average instantaneous frequencies of clean and noisy signals. Clean phonemes are convolved with room impulse responses simulated using the Image-Source Method (ISM) [1] to match the environment of 1) a livingroom, 2) a meeting room, and 3) a class room. Randomly selected noises from the RWCP sound scene database~\cite{nakamura2000acoustical}
 are added in order to simulate noisy domestic backgrounds of SNRs varying from 20dB to -15dB. The simulated microphone setup includes three microphones arranged in a 30-cm equidistant linear array located in the center of each room where a moving source is assumed to form a small spiral trajectory three meters away from the array. Overall, 100 instances of each phoneme are simulated for each of the 21 SNR-$T_{60}$ combinations, resulting in approximately 2k  signals. As  evidenced in Fig.~\ref{fig:timit}, the relative improvements gained by using MMD  increase as  conditions get more difficult, especially for low SNR values where it appears that the robustness of Teager energy in low-frequency bands~\cite{dimitriadis2009comparison} benefits vowels the most compared to nasals, stops, and fricatives in which offers modest improvements.
\section{Frequency Modulation Features}
\label{sec:features}
First order statistics over the frequency micro-modulations $f_k(t)=\omega_k(\,t)/2\pi$ have yielded improved results  combined with MFCCs in noisy LVCSR tasks~\cite{dimitriadis2015use}. Herein, the Mean Instantaneous Frequencies (MIF) are considered as the basic modulation features, which are the average instantaneous frequencies of each band $k$ over frames of $25$ms that are processed every $10$ms. MIFs are compared with 
richer descriptors for DNN acoustic modeling, such as the proposed Compressed Instantaneous Frequencies (CIF) and the bottleneck features derived from hierarchical deep networks as described in the following paragraphs.  The compared frequency modulation features are extracted after single- and multi-channel demodulation following the proposed MMD approach.

\subsection{Compressed Instantaneous Frequencies (CIF)}
\begin{figure}[t!]
\centering
\includegraphics[width=\columnwidth]{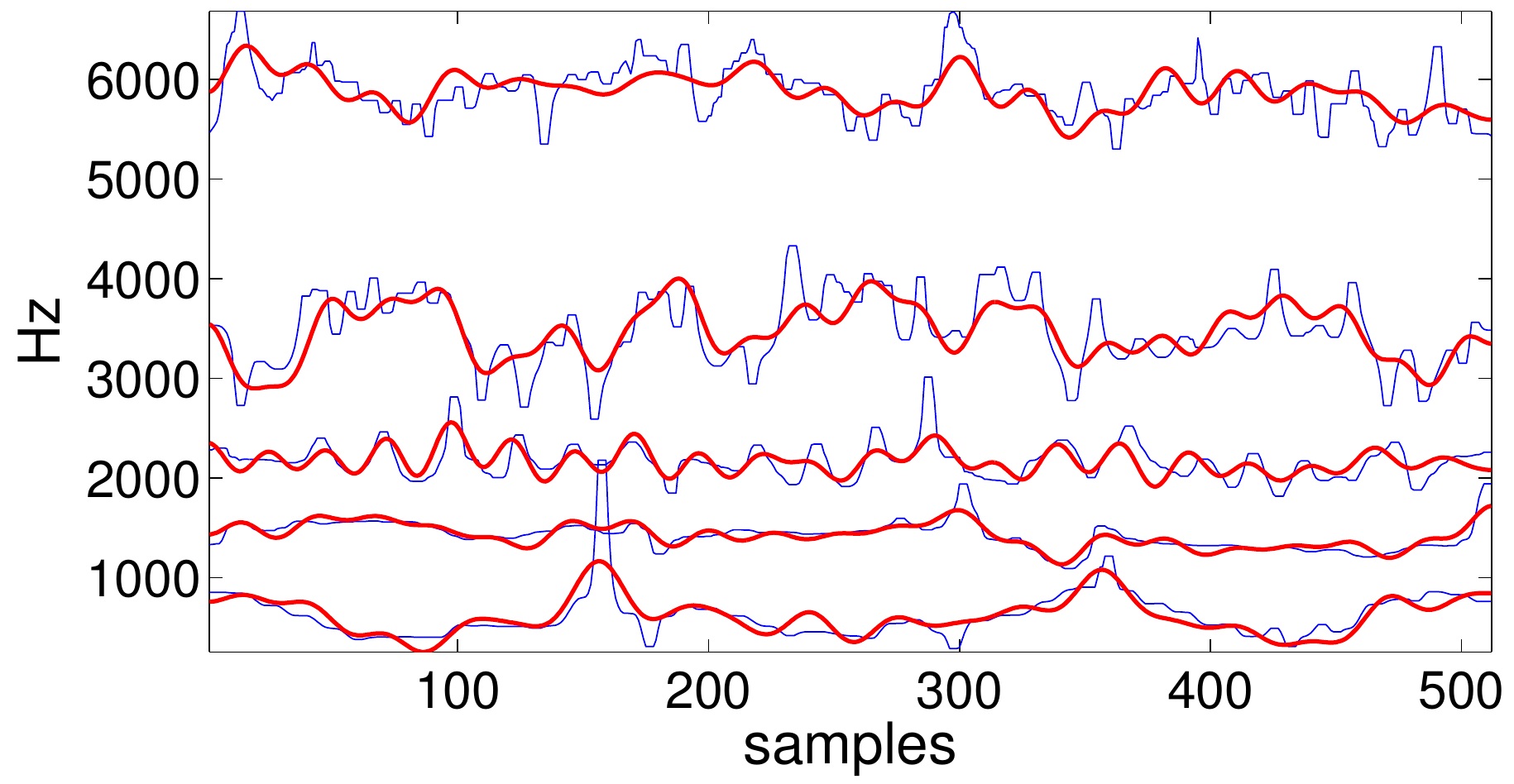}
\caption{Instantaneous frequency modulations in six Mel-spaced bands of  phoneme ``ah" and their reconstructions (red thick lines) using 10 DCT coefficients.}
\label{fig:cif}
\end{figure}

As depicted in Fig.~\ref{fig:cif}, the estimated instantaneous frequencies $f_k(t)$  contain periodic patterns that can be described compactly with a few of its Discrete Cosine Transform (DCT) coefficients. The exact number of the selected coefficients is a trade-off between the reconstruction error they achieve and their dimensionality compared to the fidelity of the employed network in which they are fed for the extraction of bottleneck features from larger temporal contexts. An example of reconstructing the instantaneous frequencies  in each band after using 10 DCT coefficients is also depicted in Fig.~\ref{fig:cif}. Generally speaking,  modulations are expected stronger and more noisy in higher bands where the filters are wider. 

\subsection{Hierarchical Deep Bottleneck Features}
\begin{figure}[t!]

\tikzstyle{block} = [draw, fill=blue!20, rectangle, 
    minimum height=1.5em, minimum width=3em,rounded corners]
\begin{tikzpicture}[auto, node distance=1cm,>=latex']
 \node [coordinate, name=in1] {};
 \node [draw, block ,node distance=2cm, right of=in1] (dnn1) {DNN};
 \node (AuxNode01) [node distance=0.5cm, below of=dnn1] {};
  \node (AuxNode02) [node distance=0.5cm, below of=in1] {};
 \node [coordinate, name=in2,below of=AuxNode02] {};
 \node [block,below of=AuxNode01] (dnn2) {DNN};
 \node [block,fill=green!20,minimum height=4em, minimum width=2em,node distance=2cm,right of=AuxNode01] (pca) {PCA}; 
 \node [block,node distance=2.2cm,right of=pca] (dnn3) {DNN}; 
  \node [coordinate, name=out, node distance=1.5cm,right of=dnn3] {};

 \draw [draw,->] (in1) -- node[left,text width=4em] {\small MFCCs} node[above] {\small $9 \times 12$} (dnn1);
  \draw [draw,->] (in2) -- node[left,text width=4em] {\small MIFs} node[above] {\small $9 \times 12$} (dnn2);
 \draw [draw,->] (dnn1) -- node {\small $42$} (pca);
 \draw [draw,->] (dnn2) -- node {\small $42$} (pca);
  \draw [draw,->] (pca) -- node {\small $9\times 42$} (dnn3);
    \draw [draw,->] (dnn3) -- node {\small $42$} (out);

\end{tikzpicture}
\caption{Extraction of 42 deep hierarchical bottleneck features after transforming and combining MFCCs  with mean instantaneous modulation frequencies  (MIFs) spanning  contexts of approximately 800ms ($9\times9=81$ frames).}
\label{fig:diagram}
\end{figure}
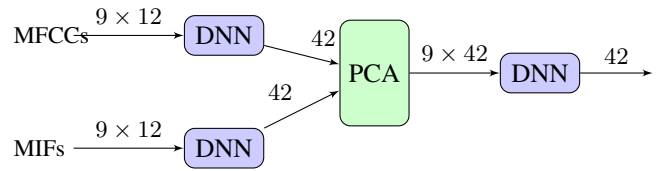

The complementary MFCC and frequency modulation features are transformed and combined through a hierarchical network of bottleneck DNNs for the extraction of long-term deep features which in turn are augmented with speaker adapted features. As shown in Fig.~\ref{fig:diagram}, first, compression of 9-frame temporal contexts is realized for each feature set through bottleneck networks. Subsequently, the activations of their bottleneck layers are concatenated and given in 9-frame vectors to the combination network after reducing their dimensionality by applying Principal Component Analysis (PCA), retaining 95\% of the total variability.
The final feature vector is formed after augmenting the bottleneck features  of the combination network with the initial MFCCs transformed using  feature-space Maximum Likelihood Linear Regression (fMLLR).

\section{Experimental Framework}
\label{sec:experiments}

\subsection{Mutli-microphone DSR corpora}
\label{sec:data}
\paragraph*{DIRHA-English corpus}
The  corpus~\cite{ravanelli2015dirha} includes one-minute sequences simulating real-life scenarios of voice-based domestic control. Real far-field speech was recorded in a Kitchen-Livingroom space by 21 condenser microphones arranged on pairs and triplets on the walls, and pentagon arrays on the ceilings. 12 US and 12 UK English native speakers were recorded on WSJ, phonetically-rich, and home automation sentences. Moreover, clean speech was recorded in a studio by the same speakers, on the same material, and convolved with the corresponding room impulse responses to produce simulated far-field speech.  Overall, 1000 noisy and reverberant utterances of real (dirha-real) and simulated (dirha-sim) far-field multichannel speech  were extracted by the sequences and used for experimentation. In our experiments, beamforming is applied on the six channels (LA1-LA6) of the pentagon ceiling-array located in the livingroom.


\paragraph*{AMI corpus}
The proposed features are also tested on the three tasks of the AMI meeting corpus~\cite{Car+06}
which consists of 100 hours  of meeting recordings
captured, transcribed and  
organized for DSR benchmarking according to three microphone setups:
 a) individual headset microphones (IHM), b) single distant microphone (SDM), and c) multiple distant microphones (MDM).  
 The three  tasks offer us the opportunity to test the robustness of the proposed features on various  setups. For the MDM scenario, the eight channels of the 10cm radius circular table array are combined via beamforming. Overlapping speech segments are excluded from our experiments. Additionally, the employed trigram language model is trained only on the transcriptions of the \textit{train} set, without using the Fisher transcriptions as the standard Kaldi recipe supports. We report results on the \textit{eval} set.

\paragraph*{CHIME-4 corpus}
The CHiME-4 task~\cite{vincent2017analysis} is a far-field speech recognition challenge for single- and multi-microphone tablet device recordings in everyday scenarios under four noisy environments: street (STR), pedestrian area (PED), cafe (CAF) and bus (BUS). For training, 1600 utterances were recorded in the four environments from four speakers, and additional 7138 noisy utterances were simulated from WSJ0 by adding noises from the four noisy environments.
The challenge setup consists of three tracks in which recognition is realized by using one (1ch), two (2ch), or six (6ch) channels from the tablet array. Multichannel recognition (2ch, 6ch) is based on beamforming.
We report results for the three tracks on the 2640 utterances of the evaluation set, which consists of 330 utterances in each of the same eight conditions. 
Our recognition setup, as described in the following paragraphs, is based on the latest baseline Kaldi recipe in which TDNN acoustic models are trained on beamformed signals, while no RNNLM rescoring is applied.

\subsection{Feature extraction configuration}

\begin{figure*}
\centering
\includegraphics[width=0.8\linewidth]{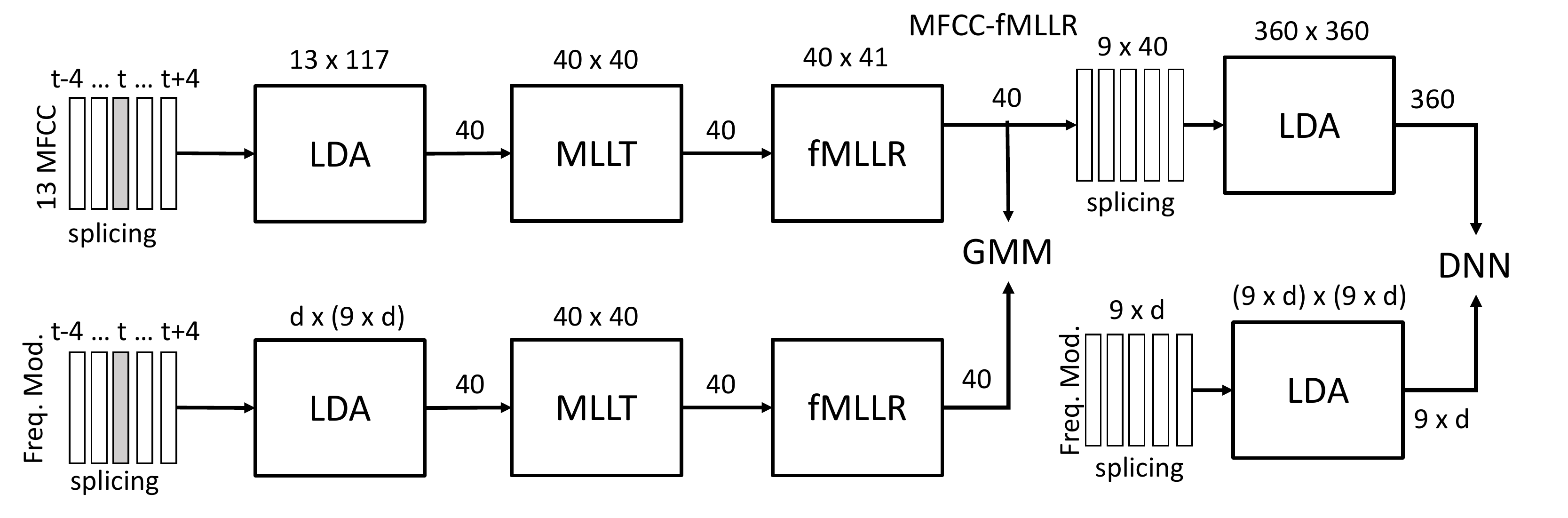}
\caption{Extraction and combination of MFCC-fMLLR~\cite{rath2013improved} features with MIF ($d=12$) and CIF ($d=60$) frequency modulation features  for GMM and DNN acoustic modeling.}
\label{fig:features}
\end{figure*}

Multiband speech demodulation is realized with a Mel-spaced filterbank of 12 and 6 Gabor filters for the extraction of 12 MIF and 60 CIF (10 DCT coefficients per filter) features for each frame, respectively. For better formant localization, the filters are overlapped by 70\% and 50\%, respectively. Instantaneous frequencies are smoothed with a 7-sample median filter in order to eliminate possible singularities that are caused by instabilities of the Teager-Kaiser energy operator in small amplitude values. 
Features are mean and variance normalized to cope with long-term effects. Standardization is applied per filter in utterance level before extracting the features in frames.
Multichannel demodulation is realized by using the same channels which are employed for beamforming according to the setup of each database.
Finally, modulation features are spliced in the same way as MFCCs and both sets are concatenated to the input of the employed networks. Note that LDA and fMMLR transformations, where they are referred, are applied separately on top of the two feature streams, as depicted in Fig.~\ref{fig:features}.

\subsection{Beamforming and data augmentation}
Speech denoising is also used in the front-end stage, in which the  available multichannel data are beamformed by using the BeamformIt tool~\cite{Anguera2007} based on the setup of each database, as described in \ref{sec:data}. The BeamformIt tool a state-of-the-art delay-and-sum beamformer that is extensively used in several  multichannel DSR systems and supports blind reference-channel selection and two-step time delay of arrival Viterbi postprocessing. 
In the absence of sufficient training data for environments with distant microphones, a practical and widely used  approach for acoustic modeling is to generate artificial training data by simulating the expected acoustic conditions of the target environment. The simulation process involves convolution of studio-quality speech with room impulse responses and noise addition in several SNR levels. We follow a slightly different approach for the case of the DIRHA-English database, where in order to increase robustness and reduce the training-testing mismatch, we generate beamformed signals for training, like the ones we intent to recognize. Thus, the ceiling-array recordings for beamforming are simulated by using RIRs measured from various positions in the room.  

\subsection{Recognition schemes}
\subsubsection{Baseline GMM-HMM System}
A baseline GMM-HMM recognizer is built  based on the standard Kaldi recipe. First, tied-state triphones are trained on 13 MFCCs with their first- and second-order derivatives and then, LDA, MLLT and fMLLR transformations~\cite{rath2013improved} are applied to train speaker independent models (\textit{tri6}). Gaussian subspace acoustic models (\textit{sgmm}) are also developed in which the universal background model (UBM) is trained on the \textit{tri6} GMMs.
Regarding language modeling, trigrams are trained on the transcriptions of the training sets.
 
\subsubsection{Tandem Recognition}
A GMM-HMM system is trained on top of the deep bottleneck features extracted by the proposed hierarchical scheme of bottleneck DNNs that is developed using TensorFlow. Each DNN consists of $6$ hidden layers ($[1048, 1048, 1048, 42, 1048, 5000]$) of tanh nonlinearities. The bottleneck layer has $42$ nodes while the last hidden layer acts like mixture components of the pdf in the softmax layer, comparable to GMMs. Nine frames are spliced and given to the input of the network which is trained to classify $2.5M$ frames to one of each $3405$ nodes of the softmax layer corresponding to the senones of the baseline GMM-HMM system that provides the frame-state alignments.  The network weights are trained layer-wise in 20 epochs by following the iterative stochastic gradient descent  training using minibatches of 256 vectors. To prevent overfitting and for adjusting the learning rate parameter, 10\% of the training corpus (chosen randomly) is used as cross-validation set.

\subsubsection{Hybrid Recognition}
Neural networks are trained  to provide pseudo log-likelihood scores for HMM decoding. Herein, DNNs~\cite{vesely2013sequence} and Time Delay Neural Networks (TDNNs)~\cite{peddinti2015time} are considered on spliced frames of MFCCs appended with modulation features. Substantially, their first layers act as feature transformation and fusion units on the combined feature sets similarly to the already described bottleneck networks. DNNs of six fully-connected hidden layers of $2048$ sigmoid nonlinearities are trained on mini-batches of $512$ samples in which 9-frame splices of $40$ fMLLR-transformed MFCCs are included. Training is realized in three stages: 1) DBN pre-training, 2) frame cross-entropy training, and 3) sequence-training optimizing the sequential Minimum Bayes Risk criterion. 
The developed TDNNs, capable  of tackling long-term interactions between speech and corrupting sources in reverberant environments, consist of five time-delay layers modeling multi-scale contexts of $\{[-2,2],[-1,1],[-1,1],[-3,3],[-6,0]\}$ compared to the running frame in time $t$. Their input features are 11-frame
 splices of 40-dimensional hi-resolution MFCCs appended with 100-dimensional i-vectors extracted using a 512-Gaussian UBM. The training data are augmented by applying 3-way speed perturbation using factors of $[0.9, 1.0 ,1.1]$ and rate perturbations by picking uniformly random values in $[0.125,2]$. 


\section{Results}
\label{sec:results}

\begin{table}[!t]
\caption{GMM-HMM recognition WERs (\%) with triphones on combinations  (``+")  of MFCC and modulation features extracted after MMD (``\_mmd") or beamforming (``\_dsb").}
\label{tab:baseline_results}
\centering
\setlength\tabcolsep{2pt}
\small
\begin{tabular}{|c||c|c||c||c|c|}
\hline
	DIRHA & MFCC &	 +MIF &	 +MIF\_mmd	& MFCC\_dsb	& +MIF\_dsb\\\hline
dirha-sim	& 62.9	&47.7	&45.1	&36.8	&37.2\\\hline
dirha-real	&67.9	&52.9	&51.6	&40.5	&38.8\\\hline
average	&65.4	&50.3	&48.4	&38.7	&\bf{38.0}\\\hline
\end{tabular}
\end{table}

\begin{table}[!t]
\caption{GMM-HMM recognition  WERs (\%) after $f$MLLR-based Speaker Adaptive Training (SAT).}
\label{baseline_results}
\centering
\setlength\tabcolsep{2pt}
\small
\begin{tabular}{|c||c||c|c|c|c|}
\hline
DIRHA & features & mono & tri & -LDA-MLLT & -fMLLR \\\hline
\multirow{2}{*}{dirha-sim} & MFCC\_dsb & 57.8 & 36.8 & 31.8 & 24.3 \\\cline{2-6}
 & + MIF\_dsb & 54.7	& 37.2 & 32.8 &	26.6 \\\hline
\multirow{2}{*}{dirha-real} & MFCC\_dsb & 61.5 & 40.5 & 30.9 & 29.5 \\\cline{2-6}
 & + MIF\_dsb & 52.3	& 38.8 & 33.4 & 31.2\\\hline
 \multirow{2}{*}{average} & MFCC\_dsb & 59.7 & 38.7 & 31.4 & \bf{26.9} \\\cline{2-6}
 & + MIF\_dsb & 53.5	& 38.0	& 33.1 &	28.9  \\\hline
\end{tabular}
\end{table}

\begin{table}[!t]
\caption{Tandem recognition WERs (\%) with  Subspace GMMs on hierarchical DNN bottleneck features appended with $f$MLLR-transformed MFCCs.}
\label{tandem_results}
\centering
\setlength\tabcolsep{2pt}
\small
\begin{tabular}{|c||c||c|c|c|c|}
\hline
DIRHA & MFCC\_dsb & +MIF\_dsb & +MIF\_mmd  & +CIF\_mmd \\\hline
dirha-sim & 23.4	& 22.8	& 22.3	& 21.6 \\\hline
 dirha-real & 29.1	& 28.8	& 28.5	& 27.8 \\\hline
average & 26.25	& 25.8	& 25.4	& \bf{24.7} \\\hline
\end{tabular}
\end{table}

\begin{table}[!t]
\caption{Hybrid recognition WERs (\%) using DNN acoustic models trained on multiple-frame combined features}
\label{tab:hybrid_results}
\centering
\setlength\tabcolsep{2pt}
\small
\begin{tabular}{|c||c|c|c|c|}
\hline
DIRHA & MFCC\_dsb-fmllr & MIF\_dsb & MIF\_mmd & +CIF\_mmd \\\hline
dirha-sim &	19.0	& 18.8 & 18.3  & 18.0\\\hline
dirha-real	& 25.0	& 24.6 & 24.3	& 23.9\\\hline
average	& 22.0	&  21.7 & 21.3	& \bf{20.9} \\\hline
\end{tabular}
\end{table}

\begin{table}[!t]
\caption{WERs (\%) on AMI corpus using xent-regularized TDNN with cleaned data and separate alignments per task. }
\label{tab:ami_results}
\centering
\setlength\tabcolsep{2pt}
\small
\begin{tabular}{|c||c||c|c|c|c|}
\hline
AMI	 & MFCC{\tiny\_dsb}+ivector	 & +MIF{\_dsb}	 & +MIF{\_mmd}	 & +CIF{\_mmd}\\\hline
IHM	& 25.7	& 25.8	& 25.8	& \bf{25.6}\\\hline
SDM	 & 50.1	& 48.2	& 48.2	& \bf{46.8}\\\hline
MDM & 43.9 & 41.1 & 40.9	& \bf{40.3}\\\hline
\end{tabular}
\end{table}

\begin{table}[!t]
\caption{WERs (\%) on CHIME-4 sim/real test sets following the baseline Kaldi recipe for TDNNs on delay-and-sum beamformed signals without using RNNLM rescoring.}
\label{tab:chime4_results}
\centering
\setlength\tabcolsep{2pt}
\small
\begin{tabular}{|c||c|c|c|c|c|c|c|c|c|c|}
\hline
CHIME-4 & \multicolumn{2}{c|}{MFCC{\tiny\_dsb}+ivector}	 & \multicolumn{2}{c|}{+MIF{\_dsb}}	 & \multicolumn{2}{c|}{+MIF{\_mmd}}& \multicolumn{2}{c|}{+CIF{\_mmd}}\\\hline\hline
 Track   & sim & real & sim & real & sim & real & sim & real \\\hline
1ch & 16.6 &	16.4 &	15.9	& 16.3	& 15.9	& 16.3	& 15.5	& 15.8\\\hline
2ch & 13.2 &	13.5 &	12.9	& 13.3	& 13.1	& 13.4	& 12.3	& 12.9\\\hline
6ch & 10.3 &	9.7 & 10.1 &	9.4 &	9.8 &	9.2 &	9.3 &	9.1\\\hline
\end{tabular}
\end{table}

We evaluate the combined feature sets on  three pipelines:\\1) GMM-HMM recognition with triphones and speaker adaptive training, 2) tandem recognition with subspace GMMs on hierarchical deep bottleneck features, and 3) hybrid recognition with DNN and TDNN acoustic models. 
Baseline recognition results on the DIRHA-English corpus are presented in Table~\ref{tab:baseline_results} where is evident how MIFs benefit MFCCs mostly a) when extracting the features from a single channel,  b) after using multichannel demodulation, and c) after beamforming that yields the lowest WER. On the other hand, as shown in Table~\ref{baseline_results}, linear transformations (LDA, MLLT and fMLLR)  deteriorates the performance because  the combined features are not uncorrelated and Gaussian like MFCCs. However, better combinations are accomplished after using DNN-based non-linear transformations.  As shown in Table~\ref{tandem_results}, hierarchical deep bottleneck features  with subspace GMMs  yield significantly better results over the SAT system. Additionally, the contribution of modulation features is increased after applying multichannel demodulation compared to beamforming. In hybrid recognition results of Table \ref{tab:hybrid_results}, the proposed features achieve modest improvements on MFCC-fMLLR for DNNs. Accordingly, they also benefit hi-resolution MFCCs with i-vectors for TDNNs,  yielding relative improvements up to 15\% over the baseline Kaldi recipes, as Tables \ref{tab:ami_results}, and \ref{tab:chime4_results} show. DSR performance is improved without degradation in clean speech as indicated by the results on the AMI IHM task.

\section{Conclusion}
\label{sec:conclusions}
A new approach is presented for robust demodulation of  the  frequency micro-modulations of speech based on multichannel speech energy tracking over the signals of a microphone array.  Better estimations of instantaneous frequencies enable the extraction of improved modulation features which are combined efficiently with standard feature sets in state-of-the-art recognition setups. Modest and consistent improvements are achieved in three challenging DSR corpora.


%

\appendices



\ifCLASSOPTIONcaptionsoff
  \newpage
\fi



\bibliographystyle{IEEEtran}
\bibliography{refs}
\end{document}